%% file: main.tex
\documentclass[letterpaper,twocolumn,10pt]{article}
\usepackage{usenix-2020-09}
\usepackage{graphicx}
\usepackage{tikz}
\usetikzlibrary{arrows.meta,positioning,calc}
\usepackage{pgfplots}
\pgfplotsset{compat=1.17}
\usepackage{enumitem}
\usepackage{longtable}
\usepackage{booktabs}
\usepackage{array}
\usepackage{tabularx}
\usepackage{caption}
\usepackage{fancyvrb}
\date{}
\title{\Large \bf When Latent Agents Lie: KV-Cache Integrity in Multi-Agent LLM Collaboration}
\author{
{\rm Luís Brito}\\
Escola Superior de Tecnologia e Gestão (ESTG),\\
Politécnico de Viana do Castelo (IPVC), Portugal
\and
{\rm Carlos Baquero}\\
Faculdade de Engenharia (FEUP),\\
Universidade do Porto (UP), Portugal
}
\begin{document}
\maketitle
\begin{abstract}

Full-KV latent memory can make multi-agent reasoning more capable, but it also turns hidden state into an integrity-critical communication object. We study a role-sequenced multi-agent protocol in which specialists send short visible commitments while transporting full KV-cache state to a coordinator. This design improves split-evidence aggregation, but it creates a failure mode that ordinary text inspection cannot see: the visible commitment can look plausible while the hidden KV state has been substituted, rescaled, or optimized.

Our evaluation shows both sides of this trade-off. On 65 transformed HiddenBench records with Qwen3-4B, full-KV latent collaboration reaches EM/F1 0.338/0.486, compared with 0.231/0.369 for matched text collaboration, with lower estimated parallel critical-path latency and about 279 MB of KV cache per example. The EM gain is directionally positive but statistically weaker; the F1 gain has stronger paired support. Qwen3-8B HiddenBench and full-validation HotPotQA runs also show latent-over-text EM/F1 gains. An influence-map diagnostic further shows that removing an essential specialist flips 10 of 21 latent-correct HiddenBench cases, indicating that some correct latent answers causally depend on specialist-specific hidden-state memory.

The same channel is fragile under malicious specialists. Targeted false commitments steer some answers, while random latent-thought corruption and scale-8 hidden-state manipulation collapse Qwen3-4B latent performance near zero. Visible verifier filtering misses this failure mode because it audits commitments, not transported KV. Magnitude quarantine handles evaluated naive nonsemantic attacks, but norm-matched adaptive attacks evade it: a HiddenBench white-box attack reduces EM/F1 from 0.323/0.471 to 0.077/0.119 without detection, and a 100-record HotPotQA port reduces 0.470/0.641 to 0.160/0.252 with 0/100 malicious states rejected. A full-65 calibrated learned sanitizer also fails to beat matched random-budget dropping under detector-aware attack.

The constructive result is therefore not latent anomaly detection, but transport-layer integrity. We implement an HMAC-SHA256 manifest that binds specialist identity, session, model, visible commitment, tensor metadata, and payload digest. It accepts 774/774 honest replayed payloads and rejects 295/295 recorded tampered payloads. After detected transport failure, fail-closed rejection reaches 0.338/0.476 on HiddenBench; on HotPotQA, drop recovers to 0.450/0.592 and visible downgrade to 0.460/0.597. The mechanism detects post-handoff transported-KV substitution under an uncompromised transport key. Compromised endpoints that sign malicious in-band KV and semantic malicious specialists remain outside its protection boundary.

\end{abstract}

\section{Introduction}
\label{sec:introduction}

Multi-agent language-model systems increasingly divide work across specialists that hold different evidence, intermediate reasoning, or tool outputs. Modern LLM serving systems also treat KV cache as a managed state object, with disaggregated serving and KV-cache-centric storage layers \cite{zhong2024distserve,qin2025mooncake}. Latent collaboration brings these trends together: a specialist can expose a short visible commitment while transporting a much richer full-KV hidden-state object to a coordinator. That object can carry useful evidence, but it is opaque to ordinary text inspection and difficult to audit when one participant is unreliable. Recent work explores cooperative latent communication and cache-level collaboration \cite{pham2023letmodelsspeakciphers,zou2025latentcollaboration,du2025interlat,dery2026kvcachealignment,fu2025cachetocache}, while separate work studies latent attacks, safe KV sharing, text-channel multi-agent attacks, and distributed-information reasoning \cite{wang2026outofsight,asif2026lcguard,lee2024promptinfection,kavathekar2025tamas,li2026hiddenbench}.

This trade-off is especially important in split-evidence tasks, where no single specialist may hold all information required for the final answer. Existing literature motivates the components of this setting while leaving open the integrity question created by transported hidden state: when does full-KV memory carry specialist-specific evidence, and how does that state fail under malicious specialists and simple audit defenses?

We study a planner-specialist-verifier-coordinator architecture. Specialists operate in parallel, emit compact visible commitments for audit, and pass full-KV latent memory to the final coordinator. We focus compromise on specialist roles, matching settings where intermediate evidence holders can be compromised while the final coordinator remains fixed. We separate endpoint/specialist attacks, where a compromised specialist can emit malicious KV, from on-path transport tampering after honest KV has been produced. We call the resulting problem empirical KV-transport integrity: whether visible auditing, latent-state diagnostics, and transport checks can detect or reduce harmful changes to the hidden object consumed by the coordinator. The primary attack and defense experiments use \texttt{Qwen/Qwen3-4B} on the 65-record transformed official HiddenBench set with greedy decoding. Additional experiments add Qwen3-8B HiddenBench replication, full-validation HotPotQA diagnostics for clean utility and naive nonsemantic attacks, and a 100-record HotPotQA white-box/provenance slice. The scope is high-bandwidth full-KV memory; compact latent bridges inform limitations.

The main empirical result is a concrete capability/risk trade-off. In clean runs, full-KV latent collaboration improves observed EM/F1 over matched text in every completed clean evaluation; Qwen3-8B HiddenBench and full HotPotQA provide the strongest clean support, and a full-HiddenBench influence map shows that removing an essential specialist flips 10 of 21 latent-correct diagnostic answers. The security story then escalates as an integrity audit: targeted-false influence traces some steering to malicious KV memory, random and scale-8 corruptions show that transported state can dominate the coordinator, visible filtering shows that inspecting commitments does not validate hidden state, and adaptive endpoint attacks expose the boundary of magnitude quarantine. A full-65 calibrated learned sanitizer sets the anomaly-detection boundary because it trails matched random-budget drop under detector-aware attack. The constructive boundary is on-path transport integrity: an HMAC-SHA256 manifest gate accepts 774/774 honest replayed payloads and rejects 295/295 recorded tampered payloads. After detected transport failure, fail-closed drop restores damaging post-handoff substitutions on HiddenBench, and the 100-record HotPotQA port shows that adaptive hidden-state damage and provenance recovery are not unique to HiddenBench. This MAC boundary covers post-handoff transported-KV substitution under an uncompromised transport key; compromised endpoints, remote attestation failure, and semantic malicious specialists remain outside its protection boundary.

The paper makes five empirical contributions:

\begin{enumerate}[leftmargin=*]
\item We formulate role-sequenced full-KV latent collaboration as an empirical transported-state integrity audit.
\item We evaluate high-bandwidth latent memory for split-evidence reasoning against text, local-only, voting, oracle, naive latent, and verifier-filtered latent branches, with Qwen3-8B and full transformed HotPotQA diagnostics.
\item We introduce a malicious-specialist evaluation separating semantic false-evidence steering from nonsemantic hidden-state corruption, high-norm manipulation, sign-flip controls, and colluding variants.
\item We add latent influence-map diagnostics showing clean specialist-specific evidence use and a bounded attack-side target-steering mechanism through malicious KV memory.
\item We analyze visible verifier filtering, latent-state quarantine, learned-sanitizer failure, and MAC-gated fail-closed rejection, showing where naive attacks recover, where adaptive white-box optimization breaks post-hoc filters, and where transport authentication gives a concrete positive systems boundary for on-path substitution.
\end{enumerate}

Together, these results show that high-bandwidth latent memory can improve clean split-evidence aggregation while creating a transported-KV integrity problem that visible commitments and simple post-hoc latent-state filtering do not close.

\section{Background and Related Work}
\label{sec:background-and-related-work}

This section situates the paper at the intersection of latent communication, KV-cache serving, latent-channel safety, multi-agent LLM security, Byzantine aggregation, and distributed-information reasoning. Prior work studies each component largely in isolation. The gap targeted here is their combination: split-evidence reasoning with malicious specialists, transported full-KV state, latent-state fusion or filtering, and matched text-channel comparisons.

\subsection{Latent Communication Between Agents}

Recent work establishes that LLM agents can communicate through representations richer than short natural-language messages. CIPHER replaces debate messages with embedding-space signals derived from model output distributions, motivating the broader idea that tokenized text may discard information useful for collaboration \cite{pham2023letmodelsspeakciphers}. Interlat studies direct latent-space communication using collected last-layer hidden states and receiver-side processing \cite{du2025interlat}. LatentMAS is closest to the high-bandwidth setting considered here: it uses latent thoughts and shared KV-cache working memory for cooperative multi-agent reasoning \cite{zou2025latentcollaboration}.

Related systems explore hybrid or heterogeneous variants. These include text plus hidden-state deltas \cite{tang2025statedelta}, KV-cache alignment across models \cite{dery2026kvcachealignment}, and cache-to-cache semantic transfer \cite{fu2025cachetocache}. Other work studies latent-variable accounts of shared and private thoughts \cite{zheng2025thoughtcommunication}. Visual-pathway latent communication addresses heterogeneous agents \cite{liu2026visionwormhole}, while hybrid latent-text communication is designed to preserve more controllability than fully latent exchange \cite{mou2026hylat}.

These papers establish latent communication as a concrete design space and delimit the present contribution. This paper shifts the question from clean latent capability to the behavior of a high-bandwidth latent channel when agents have different private evidence and some specialists are malicious. The reviewed latent-communication papers primarily address cooperative capability, efficiency, heterogeneous transfer, or hybrid protocol design; this paper studies split-evidence robustness and transported-KV integrity.

\subsection{KV-Cache Serving and Transport}

LLM serving systems increasingly expose KV cache as a managed systems object: DistServe disaggregates prefill and decoding across GPUs \cite{zhong2024distserve}, and Mooncake builds a KV-cache-centric serving architecture with disaggregated compute and cache resources \cite{qin2025mooncake}. These systems make the relevant abstraction concrete: KV state can be stored, scheduled, moved, and reused across components. That systems backdrop motivates the empirical integrity question studied here: whether transported full-KV state remains trustworthy for latent collaboration when specialist-provided hidden state can be attacked.

\subsection{Latent Attacks, KV Safety, and Leakage}

Opacity makes latent channels security-critical rather than inherently safer. The closest adversarial latent prior, Out of Sight, Not Out of Mind, studies attacks on latent-based multi-agent systems through hidden-state or KV-cache steering \cite{wang2026outofsight}. That overlap is material: it already identifies latent-only attack risk at KV handoffs, argues that visible-text inspection is insufficient, and uses controls to separate damaging latent attacks from arbitrary perturbation or invalid-generation explanations. The contribution here is the combined evaluation of role-sequenced split-evidence reasoning, malicious specialists, matched visible-text versus full-KV behavior, and state-filter diagnostics that include adaptive norm-matched breaks and transfer bounds under one protocol. THOUGHTSTEER studies backdoor attacks on continuous latent reasoning by perturbing latent vectors while evading token-level defenses \cite{parekh2026thinkingwronginsilence}; it further supports the need to audit latent channels, while leaving multi-agent KV handoffs and split-evidence role sequencing outside its scope.

Privacy and reconstruction work gives a second reason to avoid treating latent communication as inherently benign. LCGuard treats shared KV caches as latent working memory and studies representation-level transformations for safer KV sharing \cite{asif2026lcguard}. Other work reports leakage or reconstructability from embeddings and hidden activations \cite{wan2024informationleakage,liu2024eguard,nikolaou2025injective}. AgentLeak extends the concern to multi-agent systems by emphasizing that internal channels such as inter-agent messages and shared memory can contribute to privacy exposure \cite{elyagoubi2026agentleak}. These sources motivate treating hidden representations as potentially privacy-relevant; this study uses that motivation while focusing its measurements on integrity.

\subsection{Multi-Agent LLM Security}

The broader multi-agent security literature supports the malicious-specialist threat model. Prompt Infection studies prompt-injection propagation across LLM-to-LLM interactions \cite{lee2024promptinfection}. TAMAS benchmarks adversarial risks in multi-agent LLM systems \cite{kavathekar2025tamas}. MAD-Spear studies prompt-injection attacks in multi-agent debate \cite{cui2025madspear}.

Other work analyzes recurring failure modes in multi-agent LLM systems \cite{cemri2025whydofail} and limitations of embedding-based defenses \cite{zhang2026embeddingdefenses}. Related defenses and threat analyses include credibility-based adversary resistance \cite{ebrahimi2025credibility} and credit-based dynamic threat detection \cite{feng2025sentinelnet}. They also include cooperative attacks and rectification in language-space systems \cite{luo2026star} and system-level security challenges for interacting AI agents \cite{schroederdewitt2025multiagentsecurity}.

This literature motivates adversarial participants, collusion, trust weighting, and detection as realistic concerns. Its main channel assumptions, however, are usually visible text, tools, output consensus, or system-level interaction. That leaves an important methodological mismatch for latent-memory collaboration: a visible verifier may inspect commitments or summaries, but it cannot directly audit the hidden KV states consumed by a downstream coordinator. The present manuscript uses that mismatch as a core motivation for evaluating both visible-channel filtering and latent-state diagnostics.

\subsection{Byzantine Aggregation and Robust Fusion}

Byzantine-robust distributed learning provides useful vocabulary and candidate mechanisms for malicious-participant settings. Krum selects updates by neighborhood proximity under Byzantine workers \cite{blanchard2017machine}. Coordinate-wise median and trimmed mean give robust aggregation rules with statistical analyses in distributed learning \cite{yin2018byzantinerobust}. Geometric-median aggregation appears in robust federated learning \cite{pillutla2019robust}. Bulyan and later attacks illustrate both stronger defenses and the possibility that robust rules can still be circumvented \cite{elmhamdi2018hidden,baruch2019little}.

These methods provide useful design pressure for this setting, while their native objects are gradients or model updates. Latent specialist states are inference-time messages generated from semantically different evidence partitions, and an honest specialist can be an outlier precisely because it saw information that other specialists did not. For that reason, robust fusion in this manuscript is treated empirically: Byzantine aggregation motivates baselines and diagnostics, while high-bandwidth LLM KV states require direct measurement.

\subsection{Distributed-Information Reasoning}

The evaluation setting is motivated by hidden-profile and fragmented-evidence tasks. HiddenBench is the closest benchmark precedent because it focuses on collective reasoning under distributed information, where no single participant necessarily observes all decision-relevant evidence \cite{li2026hiddenbench}. Other multi-agent and retrieval-augmented systems also address fragmented information. MACT studies multi-agent table question answering with tool use \cite{zhou2025mact}. RAGentA and MASS-RAG use multi-agent retrieval and synthesis for attributed or evidence-based question answering \cite{besrour2025ragenta,xiao2026massrag}. Federated or privacy-preserving RAG systems consider knowledge distributed across silos \cite{addison2024cfedrag,gao2026fdrag,mao2026fedrag}.

These works support the premise that distributed evidence is a meaningful evaluation regime. They leave open the combined setting targeted here: high-bandwidth transported KV state, malicious latent participants, robust latent filtering or aggregation, and matched text-versus-latent attack comparisons. This specific gap motivates a controlled empirical audit of transported-KV integrity for split-evidence multi-agent reasoning under malicious specialists, with explicit comparison to text collaboration and with robustness claims tied to the reported evidence.

\section{Problem Setting and Threat Model}
\label{sec:problem-setting-and-threat-model}

We study split-evidence question answering in which the complete evidential context is distributed across specialists. A task instance contains a query, an answer key, evidence pieces, identifiers for required evidence, and agent-specific evidence partitions. In the transformed HiddenBench records used for the primary evaluation, shared information is assigned to every specialist, while hidden information is distributed across specialists and treated as structurally required because the source benchmark does not annotate minimal supporting-fact rationales. The setting therefore targets distributed-information reasoning rather than ordinary single-context question answering.

The system has four logical roles, summarized in Figure 1. The planner receives the query and records the evidence needs for the instance. Specialists receive the query, the planner state, and their assigned evidence partitions; they are the compromised role in the main threat model. Each specialist may emit a visible commitment and, in latent branches, a transported full-KV state object consumed by the coordinator. The verifier is an audit component that observes specialists' visible commitments and can filter contributions, while raw hidden-KV semantics remain outside its view. The final coordinator combines accepted information and produces the answer. This section fixes the roles and adversary scope; the next section describes the operational information flow.

We distinguish evidence ownership from answer ownership. A specialist is responsible for its evidence partition and any claim it makes about that partition. The coordinator is responsible for combining accepted specialist contributions into a final answer. The verifier can inspect visible commitments, but a plausible visible commitment is insufficient to establish that the associated latent state is benign. The integrity property we study is whether the transported hidden object remains aligned with the specialist's evidence and visible commitment well enough to support correct coordination. For post-handoff substitution, we instantiate this boundary with a transport-layer MAC over a canonical KV manifest; the broader study is an empirical audit.

The attacker controls one or more specialist agents after evidence partitioning. The attacker may try to steer the coordinator toward a target wrong answer, suppress useful evidence, or corrupt the latent state associated with a compromised specialist. The attacker does not change the benchmark record, the answer key, the honest specialists' evidence partitions, the planner, the verifier code, the coordinator code, or the evaluation metric. We use the following capability ladder; Figure 1 marks the channel each tier touches.

\begin{itemize}[leftmargin=*]
\item \textbf{Tier A: semantic-only specialist control.} The attacker controls a compromised specialist's visible text and commitment content, but cannot read, write, or perturb hidden states in transit. This tier models a malicious or unreliable evidence holder that lies through the ordinary textual interface.
\end{itemize}

\begin{itemize}[leftmargin=*]
\item \textbf{Tier B: hidden-state write access without gradients.} The attacker can overwrite or transform a compromised specialist's latent-thought KV states before the coordinator consumes them, but cannot compute gradients through the victim coordinator. This tier covers nonsemantic hidden-channel manipulations such as random state corruption, high-magnitude scaling, sign reversal, norm-matched substitutions, and transfer-style perturbations constructed without per-victim gradient optimization. The key distinction from Tier A is that the visible commitment may remain plausible while the hidden state has been changed.
\end{itemize}

\begin{itemize}[leftmargin=*]
\item \textbf{Tier C: white-box gradient access.} The attacker has Tier B write access and can also differentiate through the victim coordinator, including the model and the relevant forward computation. This tier represents a strong adaptive adversary and bounds attacks that search for damaging hidden-state directions while remaining inside the magnitude range expected from honest specialists.
\end{itemize}

Hidden-state write access is an implementation-level cache threat: latent memory is an object passed between specialist and coordinator components, and an attacker with control of a specialist runtime, serving node, or model-hosting infrastructure on that path could substitute, rescale, or perturb the KV cache before coordination. This capability is stronger than ordinary natural-language control and is therefore separated from Tier A. The evaluation keeps the planner, verifier, coordinator, honest specialists, dataset, and metrics fixed.

The threat model therefore separates two integrity boundaries. An endpoint/specialist adversary controls the specialist process that emits or signs the KV object; this is where the semantic, nonsemantic, and white-box attacks live, and a transport MAC does not make endpoint-generated malicious KV benign. A transport/on-path adversary tampers with an already-produced honest KV object after handoff; the HMAC check targets this boundary by authenticating specialist identity, record/session identity, model identity, visible-commitment hash, tensor metadata, and a SHA-256 payload digest under an uncompromised transport key.

The focus on KV substitution isolates the channel-specific integrity question created by full-KV collaboration: whether a compromised specialist can leave the visible commitment plausible while changing the hidden object consumed by the coordinator. We use "KV-transport integrity" in this empirical sense: the question is whether the coordinator's transported hidden input is trustworthy under the evaluated specialist and transport compromises. The concrete transport check detects post-handoff substitution while leaving endpoint-generated semantics to the specialist trust boundary.

White-box access is a stronger tier. It applies to research, self-hosted, or shared-model settings where an attacker can obtain the model weights and reproduce the coordinator computation, or to infrastructure compromise that exposes the differentiable serving stack. Accordingly, Tier C is reported as an adaptive upper bound for the evaluated threat model.

The scope targets specialist compromise, transported-KV substitution, and registered benchmark evidence. Separate evidence is needed for remote attestation, key-management analysis, endpoint-compromise defense, differential privacy, majority-malicious coalitions, compromised planners or coordinators, heterogeneous-model interoperability, production serving measurement, and broad benchmark generalization. Separate evidence is also needed for compact latent bridges, complete robust fusion through visible verification, adaptive-security guarantees for magnitude quarantine, and semantic malicious specialists that generate and sign their own KV in-band. Privacy and leakage are treated as literature-motivated concerns rather than measured outcomes in this study.

\section{Role-Sequenced Latent Memory Protocol}
\label{sec:role-sequenced-latent-memory-protocol}

\input{figures/fig_protocol.tex}

The evaluated system is a fixed role sequence with a planner, multiple specialists, an optional verifier, and a final coordinator; Figure 1 summarizes the roles, channels, and attack surface. The central object is the specialist-to-coordinator full-KV handoff: a transported hidden-state object that may carry useful evidence but is not directly inspectable as text. The main claims concern this role-sequenced full-KV branch; earlier compact-message harnesses inform diagnostics and limitations.

As Figure 1 shows, the planner receives the query and records the expected evidence needs, and each specialist then receives the query and its private evidence partition. Every specialist emits a compact visible commitment and, in latent branches, full key-value cache state (prompt-token KV state plus generated latent-thought KV tokens). After optional filtering, the coordinator answers from all specialist full-KV memories in the naive latent branch, only verifier-accepted full-KV memories in the filtered latent branch, or visible specialist commitments in the matched text branch. This is the high-bandwidth handoff whose empirical integrity is tested by the attack and defense suite.

The verifier sits between specialists and the coordinator. It scores the visible commitments and can remove a specialist before that specialist's text or latent contribution is used downstream, while raw KV state remains outside its inspection path. Visible commitments therefore provide an audit and filtering hook; transport integrity is handled at the KV boundary. The matched text branch keeps the same role sequence, records, and role budget, but it replaces hidden full-KV memory with visible text commitments; it is a visible-channel comparison with a different attack surface from latent-state manipulation.

The implementation logs protocol accounting separately from answer quality, including commitments, verifier decisions, KV summaries, latency, and bandwidth. Because specialist work is conceptually parallel, the efficiency comparison reports an estimated parallel specialist critical path in addition to measured sequential runtime. On the full clean run, naive full-KV latent collaboration transferred about 279 MB of KV cache per example on average; the matched text branch transferred no KV cache. The corresponding estimated parallel critical paths were 2.379 seconds for latent collaboration and 3.009 seconds for text role collaboration. These quantities characterize the cost of the protocol and should be read together with the result tables.

Earlier development harnesses are reported separately from this final protocol. The shared harness established common mode names, deterministic artifact logging, fixed-length latent-message plumbing, and smoke tests using development and small transformer backends. Those artifacts validate software pathways; model performance is evaluated in the final protocol. Likewise, the soft-prompt, pooled-vector, compact KV, and other compact-bridge diagnostics support limitation and sensitivity analysis for compact interfaces.

The same role sequence also supports defense diagnostics, which are reported separately from the protocol definition. Visible-commitment filtering is evaluated as an audit mechanism, and latent-state quarantine is evaluated as a post-hoc state-aware diagnostic that rejects anomalous specialist KV statistics before coordinator decoding. The quarantine result supports a scoped mitigation claim for the evaluated nonsemantic random and scale-8 hidden-state attacks under the recorded threshold and threat model. For the on-path transport boundary, the provenance check uses HMAC-SHA256 over a canonical KV manifest that binds the specialist, record/session, model, visible commitment, tensor metadata, and payload digest before coordinator acceptance. Accordingly, the protocol is an auditable, high-bandwidth full-KV latent-memory system with a concrete post-handoff substitution check, matched text branch, and explicit latency/KV accounting; compactness, privacy, key management, remote attestation, endpoint compromise, and general robustness require separate evidence.

\section{Benchmark and Experimental Setup}
\label{sec:benchmark-and-experimental-setup}

The primary benchmark is the transformed official HiddenBench set. The dataset package preserves the raw Hugging Face artifact retrieved on 2026-06-06, including source revision and raw-file hash, and converts each item into a split-evidence record with answer keys, evidence pieces, partitions, validity fields, provenance, and source-format notes. A transformation-validity audit verified all 65 official records: 250 shared items and 253 hidden items are preserved, each hidden item is assigned to exactly one specialist, and no specialist receives all required hidden evidence. Because the source lacks minimal supporting-fact rationales, all hidden identifiers are marked structurally required; rationales exist for 57/65 records, and shared-decoy language is recoverable in 48/57. We treat transformed HiddenBench as a structurally faithful split-evidence stress test with documented limits on original distractor-property reconstruction. HotPotQA provides second-benchmark diagnostics for clean utility, naive nonsemantic attacks, quarantine, decoding length, compact-KV exploration, and a reduced white-box/provenance slice. The full-validation HotPotQA suite ran on remote hardware; the 100-record adaptive/provenance slice used the local ROCm environment.\footnote{Remote HotPotQA diagnostic hardware: Ubuntu 24.04.3 LTS; Intel Xeon Processor (SapphireRapids), 24 logical CPUs; 117 GiB memory; 1x NVIDIA L40S GPU.}

All primary full-set clean, attack, and defense runs use \texttt{Qwen/Qwen3-4B} with deterministic greedy decoding, \texttt{max\_input\_tokens=4096}, \texttt{max\_new\_tokens=32}, hidden size 2560, and \texttt{torch.float16} weights. Result JSON files record \texttt{cuda:0}; the reproducibility manifest identifies this as a ROCm-backed PyTorch environment exposed through the \texttt{torch.cuda} namespace.\footnote{Local primary-run hardware: Ubuntu 24.04.4 LTS; AMD RYZEN AI MAX+ 395 w/ Radeon 8060S, 32 logical CPUs; 124 GiB memory; AMD Radeon 8060S Graphics GPU.} The environment snapshot records software, hardware, package, git, and Hugging Face-token-presence metadata without storing secrets.

The clean branches are local-only specialist answering, independent majority vote, full-evidence oracle answering, text role collaboration, naive full-KV latent role collaboration, and verifier-filtered full-KV latent role collaboration. Local-only is logged per specialist, yielding 253 per-agent rows; the other branches use 65 record-level outputs. Latent branches use high-bandwidth full-KV memory with five latent steps and identity-norm alignment to the average input-embedding norm, distinguishing them from compact soft-prompt, pooled-vector, or fixed-length latent-bridge experiments. Specialist work is executed sequentially for deterministic logging, while efficiency summaries also report the estimated parallel specialist critical path.

The adversarial suite keeps the same model, dataset, decoding, and role-sequenced implementation while compromising specialist roles. The full-65 core suite uses one compromised specialist per record, selected as the last sorted specialist identifier, and evaluates targeted false specialist behavior, random latent-thought KV replacement at scale 1.0, scale-8 latent-thought multiplication, and latent-thought sign flip with factor -1.0. The targeted-false attacker receives the possible answers and its private evidence and is instructed to support a wrong target. Nonsemantic hidden-state attacks modify the selected specialist's latent-thought KV tokens while leaving prompt-token KV intact. Matched text comparisons are visible-channel budget or salience analogues with a different attack surface from hidden-state attacks.

Verifier and state-filter diagnostics are logged separately from the clean and attack branches. Visible verifier filtering scores specialist commitments before accepted text or latent contributions reach the coordinator, measuring a text-side audit hook rather than direct inspection of hidden KV state. Latent-state quarantine applies a post-hoc threshold to per-specialist latent-thought KV statistics for random and scale-8 attacks, then reports accepted KV bandwidth, rejection decisions, text fallback, and quarantined latent performance. Full-validation HotPotQA quarantine/drop runs use this related state-filter diagnostic on 7,405 records for Qwen3-4B and Qwen3-8B; a separate 100-record HotPotQA run ports the white-box gradient norm-matched attack and evaluates drop/downgrade handling when damaged KV is treated as failed provenance.

The full defense diagnostics remain centered on transformed HiddenBench. The learned-sanitizer follow-up uses cross-fitted anomaly scores, calibrated reject-1 policies, and detector-aware gradient optimization. Provenance-gated runs measure downstream utility after a detected transport failure: the coordinator drops that hidden state while retaining authenticated KV from other specialists, or uses a visible-text downgrade ablation. The concrete substitution check is a transport-layer HMAC-SHA256 helper over a canonical KV manifest and payload digest; self-tests and artifact replay accept 774/774 honest replayed payloads and reject 295/295 recorded tampered payloads. This supplies the post-handoff provenance check, while HiddenBench and HotPotQA drop/downgrade runs supply downstream utility after failure. Production serving deployment, remote attestation, key management, endpoint-compromise defense, and semantic malicious-specialist detection are outside this measurement.

The primary task metrics are exact match and F1 against transformed HiddenBench answer keys. Attack logs record target-answer selection, clean-correct flips, clean-wrong repairs where available, and prediction changes. Defense logs record rejection confusion matrices, precision, recall, and honest false-positive rate. Efficiency logs record visible tokens, measured sequential latency, estimated parallel critical-path latency, transferred KV-cache bandwidth, and accepted KV bandwidth after filtering or quarantine. Matched-record summaries use paired bootstrap confidence intervals for EM/F1 deltas and McNemar tests for EM comparisons, keeping text-versus-latent and attack-versus-clean comparisons tied to the same 65 records.

Subset influence diagnostics re-encode every non-empty accepted specialist subset and compare full-set predictions to subset predictions. The clean influence diagnostic covers all 65 transformed HiddenBench records under matched text and latent conditions; the attack-causal diagnostic reuses the clean 16-record offset-40 slice and adds a targeted-false latent slice. These are accepted-subset causal diagnostics rather than literal deletion from a single fixed cache.

Reproducibility is handled through a generated run manifest and command list covering dataset transformation, smoke tests, primary clean and attack runs, quarantine and adaptive-defense diagnostics, optional Qwen3-8B and HotPotQA diagnostics, the 100-record HotPotQA white-box/provenance run, transport-MAC replay, metric aggregation, failure-case extraction, and manifest generation. The manifest links results, reports, run directories, artifact checks, metric tables, figure inputs, local ROCm metadata, and remote L40S provenance back to source artifacts.

\section{Clean Collaboration and Efficiency}
\label{sec:clean-collaboration-and-efficiency}

Across the completed clean evaluations, full-KV latent collaboration consistently improves observed EM/F1 over matched text role collaboration. The 65-record Qwen3-4B HiddenBench run provides the protocol-matched anchor for the full attack/defense suite, while Qwen3-8B HiddenBench and full transformed HotPotQA provide the strongest clean-utility support. Table 1 reports the Qwen3-4B HiddenBench clean utility, latency, bandwidth, and compute-proxy measurements for local-only specialist answers, independent majority voting, a full-evidence oracle prompt, text role collaboration, naive high-bandwidth full-KV latent collaboration, and verifier-filtered full-KV latent collaboration. The latent rows evaluate full-KV memory transport; compact latent-bridge claims are outside the result in this table.

\begin{table*}[t]
\centering
\begingroup
\scriptsize
\setlength{\tabcolsep}{3pt}
\renewcommand{\arraystretch}{1.12}
\caption{Clean collaboration and efficiency on the transformed HiddenBench clean run. The local-only row contains 253 per-agent specialist answers; the other rows contain 65 record-level system outputs.}
\begin{tabularx}{\textwidth}{@{}>{\raggedright\arraybackslash}X>{\raggedright\arraybackslash}X>{\raggedright\arraybackslash}X>{\raggedright\arraybackslash}X>{\raggedright\arraybackslash}X>{\raggedright\arraybackslash}X>{\raggedright\arraybackslash}X@{}}
\toprule
Clean mode & Rows & EM & F1 & Parallel critical path (s) & Mean KV MB & Mean compute proxy tokens \\
\midrule
Local-only agents & 253 & 0.202 & 0.345 & 0.521 & 0.00 & 449.2 \\
Independent majority & 65 & 0.185 & 0.336 & 0.538 & 0.00 & 1748.2 \\
Full-evidence oracle & 65 & 0.277 & 0.403 & 0.689 & 0.00 & 602.3 \\
Text role collaboration & 65 & 0.231 & 0.369 & 3.009 & 0.00 & 2474.7 \\
Latent full-KV role collaboration & 65 & 0.338 & 0.486 & 2.379 & 279.03 & 2278.6 \\
Latent full-KV verifier filtered & 65 & 0.308 & 0.442 & 4.470 & 146.10 & 5420.0 \\
\bottomrule
\end{tabularx}
\endgroup
\end{table*}

The strongest clean row in Table 1 is naive full-KV latent collaboration, at EM/F1 0.338/0.486, compared with 0.231/0.369 for text role collaboration on the same 65 records. A matched-record paired analysis sets the statistical scope: the latent-minus-text EM difference is +0.108 with a 95\% confidence interval of [-0.015, +0.231], and the exact McNemar p-value is 0.1435. The paired F1 interval is positive, +0.118 with 95\% confidence interval [+0.002, +0.233]. The clean result supports an observed utility advantage for this high-bandwidth latent branch, with stronger evidence for F1 than for EM.

Table 2 summarizes the completed replications that strengthen the clean utility result while preserving the high-bandwidth caveat. The 65-record Qwen3-4B HiddenBench run remains the protocol-matched attack/defense anchor, but the clean-utility claim is read across the full evidence hierarchy. On the same 65 HiddenBench records with Qwen3-8B, latent full-KV reached EM/F1 0.415/0.544 versus 0.262/0.392 for text, with paired EM/F1 deltas of +0.154/+0.153 and a McNemar p-value of 0.0309. On the full 7,405-record transformed HotPotQA distractor-validation set, latent also exceeded text for both tested model sizes: Qwen3-4B improved from 0.244/0.397 to 0.450/0.625, and Qwen3-8B improved from 0.378/0.538 to 0.528/0.692. These rows support second-benchmark clean utility; the matching HotPotQA attack and quarantine diagnostics are scoped separately in the attack and defense sections.

\begin{table*}[t]
\centering
\begingroup
\scriptsize
\setlength{\tabcolsep}{3pt}
\renewcommand{\arraystretch}{1.12}
\caption{Completed clean replications beyond the primary Qwen3-4B HiddenBench run.}
\begin{tabularx}{\textwidth}{@{}>{\raggedright\arraybackslash}X>{\raggedright\arraybackslash}X>{\raggedright\arraybackslash}X>{\raggedright\arraybackslash}X>{\raggedright\arraybackslash}X>{\raggedright\arraybackslash}X>{\raggedright\arraybackslash}X@{}}
\toprule
Setting & n & Text EM/F1 & Latent EM/F1 & Delta EM/F1 & Text/latent critical path (s) & Mean KV MB \\
\midrule
HiddenBench, Qwen3-8B & 65 & 0.262/0.392 & 0.415/0.544 & +0.154/+0.153 & 6.384/3.349 & 279.03 \\
HotPotQA, Qwen3-4B & 7405 & 0.244/0.397 & 0.450/0.625 & +0.206/+0.228 & 1.104/0.355 & 133.74 \\
HotPotQA, Qwen3-8B & 7405 & 0.378/0.538 & 0.528/0.692 & +0.150/+0.154 & 1.114/0.358 & 133.74 \\
\bottomrule
\end{tabularx}
\endgroup
\end{table*}

A separate influence-map diagnostic asks whether the clean latent gain is confined to aggregate averages or whether final answers are sensitive to particular specialists' hidden-state memories. For each transformed HiddenBench record, the diagnostic enumerates every non-empty specialist subset under matched text-clean and latent-clean conditions, then computes leave-one-out and exact subset influence scores. In the full-specialist diagnostic condition, latent full-KV again exceeded text, reaching EM/F1 0.323/0.463 versus 0.246/0.384. More importantly, among latent-correct outputs, 10 of 21 had at least one essential specialist: removing that specialist from the accepted subset flipped the answer from correct to wrong. The same test found essential specialists in 7 of 16 text-correct outputs. Mean best-agent leave-one-out drops were also larger for latent, 0.154 EM and 0.138 F1, than for text, 0.108 EM and 0.127 F1. This is an accepted-subset diagnostic rather than a literal deletion from a fixed cache, and it supports the interpretation that full-KV latent memory can carry specialist-specific hidden evidence used by the coordinator.

The baseline rows put accuracy in context. Local-only specialist answers and independent majority voting have limited evidence coverage, and the full-evidence oracle prompt remains below the naive latent full-KV branch in this run. The verifier-filtered latent branch reduces accepted KV bandwidth from 279.03 MB to 146.10 MB, but it also lowers clean utility and increases the estimated parallel critical path to 4.470 s. In the clean setting, visible verifier filtering is therefore a cost and audit trade-off rather than a clean-utility improvement.

The efficiency result is a latency/bandwidth trade-off. Under the intended parallel-specialist accounting, naive latent full-KV collaboration has a lower estimated critical path than text role collaboration, 2.379 s versus 3.009 s. That latency direction comes with a large communication cost: the text branch transfers no KV cache, whereas the latent branch transfers an average of 279.03 MB of full-KV memory. The token-level compute proxy is similar in scale for text and naive latent collaboration, and the full prompt plus latent-thought KV state remains the dominant communication cost. The appropriate interpretation is a latency/KV trade-off; low-bandwidth efficiency remains a separate goal.

An optional deterministic rerun check repeated the text role collaboration and naive full-KV latent branches on the same 65-record Qwen3-4B HiddenBench setting. Across the primary run and two additional adjacent-seed runs, text remained at 0.231/0.369, latent full-KV remained at 0.338/0.486, the latent-minus-text deltas stayed at +0.108 EM and +0.118 F1, and the matched prediction-change counts versus the primary run were zero. Because all runs used deterministic greedy decoding with no sampling or dropout, this is a backend and deterministic rerun stability check; stochastic seed robustness remains a separate evaluation. The local-only, oracle, majority, and verifier-filtered rows are supplied by the primary clean baseline.

\section{Latent Attacks Under Malicious Specialists}
\label{sec:latent-attacks-under-malicious-specialists}

The malicious-specialist suite shows that the hidden-state channel can dominate the coordinator even when the visible role sequence remains fixed. Direct latent-thought corruption and high-norm manipulation collapse the latent branch in the full Qwen3-4B HiddenBench run, while semantic false specialists mainly produce target steering with moderate degradation. The full 65-record attack suite evaluates one compromised specialist per record; each attack script includes a same-run clean latent branch, which reached EM/F1 0.323/0.471 and is used for matched attack deltas. Table 3 separates semantic target steering, catastrophic nonsemantic collapse, and sign flip as a contrast condition that preserves magnitude while reversing direction.

The matched text attacked column in Table 3 is a visible-channel analogue under the same role and malicious-specialist budget. The two branches expose different attack surfaces: the text branch exposes corrupted or adversarial visible commitments, whereas the latent branch directly manipulates the compromised specialist's latent-thought KV state.

\begin{table*}[t]
\centering
\begingroup
\scriptsize
\setlength{\tabcolsep}{3pt}
\renewcommand{\arraystretch}{1.12}
\caption{Full 65-record malicious-specialist attack results. Same-run clean latent EM/F1 is 0.323/0.471; matched text attacked rows are visible-channel analogues of the latent attack budget. Target selection for the targeted-false specialist is reported in the prose because it is not defined for nonsemantic latent-state attacks.}
\begin{tabularx}{\textwidth}{@{}>{\raggedright\arraybackslash}X>{\raggedright\arraybackslash}X>{\raggedright\arraybackslash}X>{\raggedright\arraybackslash}X>{\raggedright\arraybackslash}X>{\raggedright\arraybackslash}X@{}}
\toprule
Attack family & Latent clean EM/F1 & Latent attacked EM/F1 & Latent delta EM/F1 & Clean-correct latent flips & Matched text attacked EM/F1 \\
\midrule
Targeted false specialist & 0.323/0.471 & 0.231/0.386 & -0.092/-0.085 & 9/21 & 0.323/0.456 \\
Random latent-thought corruption & 0.323/0.471 & 0.000/0.006 & -0.323/-0.465 & 21/21 & 0.292/0.436 \\
Scale-8 latent-thought manipulation & 0.323/0.471 & 0.000/0.000 & -0.323/-0.471 & 21/21 & 0.308/0.451 \\
Sign-flip latent-thought direction reversal & 0.323/0.471 & 0.354/0.476 & +0.031/+0.005 & 2/21 & 0.277/0.420 \\
\bottomrule
\end{tabularx}
\endgroup
\end{table*}

The semantic targeted-false attack steers the coordinator toward an explicit wrong answer more clearly than it produces a decisive aggregate collapse. The attacked latent branch selected the attacker target on 27/65 examples and flipped 9/21 clean-correct latent answers; the paired full-set delta is -0.092 EM with 95\% CI [-0.200, +0.015], -0.085 F1 with 95\% CI [-0.185, +0.015], and McNemar p=0.146. This supports a bounded target-steering result with moderate degradation.

An attack-causal influence diagnostic shows that this steering can be mediated by the malicious specialist's hidden-state memory. On the 16-record offset-40 targeted-false slice, the targeted latent branch reached EM/F1 0.250/0.365 versus 0.375/0.490 for the reused clean latent slice. Two of six clean-correct cases flipped to wrong, and both flipped cases selected the attack target. The broad harmful-influence signal remained below threshold: malicious negative leave-one-out rate was 0.188 and mean malicious leave-one-out EM/F1 deltas were -0.062/-0.052. The target-steering signal was stronger: full target-hit rate was 0.312, mean malicious target-hit Shapley was 0.297, and positive malicious target-hit Shapley appeared on 10/16 records. This sharpens the mechanism claim while keeping universal semantic collapse outside scope.

The nonsemantic attacks behave differently (Figure 2). Random replacement of the compromised specialist's latent-thought KV tokens and scale-8 latent-thought manipulation both reduce the attacked latent branch to zero or near-zero utility and flip every clean-correct latent answer in the full-set run. Their paired EM deltas are both -0.323 with confidence intervals excluding zero, their F1 deltas are -0.465 and -0.471, and both exact McNemar tests report p=9.537e-07. The important contrast is that these attacks corrupt the hidden latent-thought channel directly, without supplying a semantically plausible wrong answer.

A selected Qwen3-8B replication preserves the same qualitative ordering for the strongest nonsemantic attacks while showing that severity can vary by model and attack. With same-run clean latent EM/F1 at 0.415/0.544, random latent-thought corruption degraded Qwen3-8B to 0.200/0.316, while scale-8 latent-thought manipulation collapsed it to 0.000/0.000. The text analogues were weaker than the latent hidden-state attacks. This supports cross-model concern for nonsemantic hidden-state manipulation. Semantic false-specialist attacks, sign flip, adaptive attacks, and stochastic decoding remain outside the selected 8B replication.

HotPotQA now provides second-benchmark evidence for both naive and adaptive hidden-state damage. On the full transformed validation set, Qwen3-4B random latent-thought corruption collapsed clean EM/F1 0.448/0.621 to 0.000/0.001 in both seeds, and scale-8 latent-thought manipulation reached 0.000/0.000. On Qwen3-8B, random corruption was less catastrophic but still damaging, reducing clean 0.523/0.688 to 0.366/0.498 and 0.362/0.491, while scale-8 again collapsed the latent branch. A reduced 100-record Qwen3-4B HotPotQA port of the white-box norm-matched attack reduced clean latent EM/F1 from 0.470/0.641 to 0.160/0.252 while the magnitude filter rejected 0/100 malicious specialists. This closes much of the single-benchmark adaptive-risk objection, while leaving semantic HotPotQA target steering and full 7,405-record adaptive HotPotQA evaluation open.

Sign flip is the contrast condition for this attack family. It reverses the compromised specialist's latent-thought KV direction while preserving the attack's latent-thought scope and approximate magnitude, yet the attacked latent branch remains near the same-run clean baseline and flips 2/21 clean-correct answers. Its paired uncertainty spans zero (+0.031 EM with 95\% CI [-0.046, +0.108]; +0.005 F1 with 95\% CI [-0.059, +0.069]; McNemar p=0.688). The registered attacks indicate that catastrophic failure in this setup is tied to random replacement or high-norm manipulation of latent-thought state, not arbitrary latent perturbation.

Several boundaries frame these results. The main attack suite uses deterministic greedy decoding, one transformed 65-record HiddenBench set, and one compromised specialist in the full-set core attacks. The Qwen3-8B evidence is a selected replication for the strongest nonsemantic attacks, and the HotPotQA evidence combines full-validation naive random/scale diagnostics with a reduced 100-record white-box slice. Earlier 16-record diagnostics show that targeted-false behavior is slice-dependent: one slice degraded latent EM/F1 to 0.188/0.302, while another reached 0.375/0.531; both selected the target on 5/16 records. The attack-causal influence diagnostic is also a 16-record HiddenBench slice, so it supports a target-steering mechanism while broad semantic-attack generality remains open. Collusion evidence is limited to a 16-record targeted-false diagnostic: two colluding compromised specialists reached latent EM/F1 0.250/0.365 versus matched clean latent 0.375/0.490, selected the target on 5/16 examples, and lacked a monotonic degradation curve relative to the one-specialist slice. These boundaries frame the attack claim as a bounded empirical audit of high-bandwidth full-KV latent memory under malicious specialists.

\section{Defense Diagnostics and Trade-Off Analysis}
\label{sec:defense-diagnostics-and-trade-off-analysis}

\input{figures/fig_attack_defense.tex}

Post-handoff KV substitution is best handled as a transport-layer integrity problem when the adversary tampers after an honest specialist emits KV. State-aware diagnostics recover utility when endpoint/specialist attacks create detectable anomalies; adaptive endpoint attacks define the ceiling for both magnitude filtering and the calibrated learned sanitizer. By contrast, the MAC-backed provenance check detects on-path transported-KV substitution before coordinator decoding, and the HiddenBench and HotPotQA policy results show how verification failure can drive fail-closed evaluation. Figure 2 summarizes the main exact-match defense results.

\subsection{Visible Filtering}

Visible verifier filtering provides an audit hook with clear robustness limits. On the full 65-record HiddenBench suite, filtering worsened the targeted-false latent branch from EM/F1 0.231/0.386 to 0.200/0.352 while rejecting many honest commitments (TP=33, FP=89, FN=32, TN=99). Under random latent-thought corruption and scale-8 manipulation, plausible commitments masked corrupted KV; filtering reached 0.154/0.208 and 0.138/0.192 EM/F1, far below the matched clean latent branch at 0.323/0.471. The filter audits text, not transported KV.

\subsection{Magnitude Quarantine}

Magnitude quarantine addresses a specific setting: one malicious specialist, nonsemantic random or scale-8 latent-thought corruption, full HiddenBench evaluation, and a post-hoc post-key magnitude threshold. The rule rejects at most one specialist when its peer-median ratio exceeds 1.25, then rebuilds the latent path from accepted specialists.

Within that setting, the rule rejected the corrupted specialist on 65/65 random records and 65/65 scale-8 records, with no honest rejections and no clean false positives across 253 honest states. The accepted-specialist latent path recovered from near-zero attacked utility to EM/F1 0.338/0.476 for random corruption and 0.385/0.522 for scale-8 manipulation. The Qwen3-8B replication showed the same detection pattern, recovering to 0.369/0.490 and 0.354/0.475.

The full HotPotQA diagnostics reproduce this pattern at larger scale. For both Qwen3-4B and Qwen3-8B, quarantine rejected 7405/7405 malicious specialists under random corruption and 7405/7405 under scale-8 manipulation, with no reported clean or honest false positives. Qwen3-4B recovered to EM/F1 0.433/0.593 and 0.439/0.599; Qwen3-8B recovered to 0.497/0.650 and 0.500/0.653. These are related state-quarantine/drop diagnostics; adaptive HotPotQA evidence comes from the separate 100-record white-box slice below.

Two trade-offs bound the result. First, the reject cap is part of the threat model: with two corrupted specialists, reject-1 leaves many attacks accepted, while reject-2 removes roughly half of the partitioned evidence. Second, after one-malicious quarantine, latent full-KV can outperform text fallback when enough evidence remains: on HiddenBench, quarantine reached 0.338/0.476 and 0.385/0.522 for random and scale-8 attacks, compared with text fallback at 0.246/0.384. The supported claim is a conditional one-malicious mitigation; a general robust-fusion guarantee would require additional evidence.

\subsection{Adaptive Stress Tests}

The adaptive tests ask whether the magnitude diagnostic survives an attacker that optimizes around it. Three norm-matched attacks held the malicious latent-thought key magnitude at the honest peer level, the feature used by quarantine. All three stayed below threshold on 65/65 records, with rejection recall 0.0.

Two evasions preserved utility: random-direction norm-matched corruption produced EM/F1 0.354/0.492, and coherent neighbouring-record substitution produced 0.338/0.476. The destructive random and scale-8 attacks therefore depend on magnitude inflation. The white-box gradient-optimized norm-matched attack is the stronger counterexample: it preserves the inspected feature while optimizing against the coordinator's gold-answer likelihood, collapsing clean latent EM/F1 0.323/0.471 to 0.077/0.119 while quarantine rejects 0/65 malicious states. The 100-record HotPotQA port gives the same adaptive boundary in a second benchmark slice: clean latent EM/F1 0.470/0.641 falls to 0.160/0.252, and the magnitude filter again rejects 0/100 malicious specialists.

Detector follow-ups define the anomaly-detection ceiling. At a 5\% honest false-positive budget, the best cheap statistic reached recall 0.354, and a learned L2 probe improved held-out AUC while remaining weak under low-FP thresholding. A 12-record learned top-1 sanitizer recovered F1 under baseline and detector-aware attacks, with detector-aware optimization reducing malicious top-rank recall from 8/12 to 1/12. The full-65 calibrated follow-up gives the decisive anomaly-detector result: learned reject-1 improved the non-adaptive gradient attack from EM/F1 0.123/0.146 to 0.169/0.232; under detector-aware optimization it reached 0.092/0.131, below the matched random-budget control at 0.108/0.154, with 0/65 malicious specialists rejected. Learned sanitization sets the endpoint-anomaly detection boundary; transport integrity carries the on-path substitution defense claim.

\subsection{Transport Integrity and Fail-Closed Boundary}

The stronger positive result comes from changing the systems boundary from endpoint anomaly detection to transport integrity. The implementation uses HMAC-SHA256 over a canonical KV transport manifest that binds specialist identity, record/session identity, model identity, visible-commitment hash, tensor metadata, and a SHA-256 payload digest. Its tensor self-tests accepted the original KV segment and rejected single-value, scale-8, sign-flip, wrong-session, and wrong-key changes. Artifact replay accepted 774/774 honest replayed payloads and rejected 295/295 recorded tampered payloads: 564/564 honest and 195/195 tampered on HiddenBench, and 210/210 honest and 100/100 tampered on the HotPotQA slice.

The provenance-gated policy experiments evaluate what to do after an on-path transport check fails. In the full HiddenBench run, one specialist per record is treated as failed-provenance transported KV; the coordinator keeps authenticated latent KV from the remaining specialists and either drops the failed specialist or downgrades it to visible text. Fail-closed drop reached EM/F1 0.338/0.476 on all 65 transformed HiddenBench records, while visible downgrade reached 0.185/0.333. The damaging endpoint-attack payloads are used here as recovery references for the downstream policy, not as evidence that the MAC detects endpoint-signed malicious KV: drop restores F1 from 0.006 to 0.476 for random latent-thought replacement, from 0.000 to 0.476 for scale-8 manipulation, from 0.119 to 0.476 for white-box gradient norm-matched attack, and from 0.124 to 0.476 for detector-aware gradient attack.

The HotPotQA white-box/provenance slice supports the policy conclusion and sharpens the policy comparison. After the white-box norm-matched attack reduces EM/F1 to 0.160/0.252, treating the damaged specialist KV as failed-provenance recovers to 0.450/0.592 with drop and 0.460/0.597 with visible downgrade. Thus, the cross-benchmark result is that provenance handling restores much of the utility lost after a detected transport failure; the ordering between drop and visible downgrade is benchmark- and policy-dependent.

This result establishes a transport integrity boundary. The supported claim is detection of post-handoff transported-KV substitution under an uncompromised transport key, followed by fail-closed policy evaluation. Historical attack runs stored cache-row metadata without full tensor bytes, so the replay binds recorded clean-vs-tampered identity plus metadata; byte-level detection is supported by the implementation and tensor self-tests. Compromised endpoints signing malicious in-band KV, remote attestation failures, and semantic malicious specialists remain outside this boundary; the targeted-false no-defense reference remains EM/F1 0.231/0.386.

\subsection{Transfer Bounds}

The transfer diagnostic narrows the threat model for the white-box break. When optimized perturbations were applied without per-victim gradients, damage was much smaller even though quarantine evasion remained complete: neighbouring-record transfer reached EM/F1 0.246/0.386 with intervals overlapping clean, and a universal perturbation preserved utility at 0.354/0.494. The supported interpretation is asymmetric: magnitude quarantine filters the evaluated naive attacks, but it is insufficient against per-victim white-box optimization inside the honest magnitude band; catastrophic transfer to gradient-free attackers, other models, other benchmarks, or learned attack procedures remains unobserved.

\section{Failure Modes and Limitations}
\label{sec:failure-modes-and-limitations}

The main supported channel is high-bandwidth full-KV memory. Earlier soft-prompt, pooled-vector, learned-discrete, contiguous-memory, and short latent-thought KV interfaces exposed usable signal, but they fell short of consistently replacing text collaboration. The broader compact-KV ablation was also mixed. These ablations define the claim: this paper audits a high-bandwidth latent-memory path, while low-bandwidth latent communication remains open.

The positive clean result should be read through that lens. The clean latent advantage depends on large per-record KV transfer, and the Qwen3-4B exact-match interval still crosses zero, with stronger support for F1. Qwen3-8B HiddenBench and full HotPotQA strengthen utility evidence while leaving bandwidth, auditability, logging, and redaction concerns.

The influence diagnostics add mechanism evidence while preserving those boundaries. The clean map re-encodes accepted specialist subsets and shows that some correct latent outputs depend on particular specialists' hidden-state memory; the 16-record targeted-false slice shows target-steering mediation through malicious KV memory. Both diagnostics are single-model, deterministic, and tied to transformed HiddenBench; compact communication, privacy behavior, and broad cross-benchmark generality require separate evidence.

The empirical scope is deliberately bounded. The primary attack and defense claims use \texttt{Qwen/Qwen3-4B}, deterministic greedy decoding, and 65 transformed HiddenBench records. Replications add Qwen3-8B clean/nonsemantic support, 7,405-record HotPotQA scale diagnostics for naive random and scale-8 attacks plus quarantine, and a 100-record HotPotQA white-box/provenance slice. The full-validation HotPotQA diagnostics serve as related state-quarantine/drop evidence; the reduced adaptive slice supports adaptive-risk and provenance recovery without claiming semantic HotPotQA behavior or full 7,405-record adaptive coverage.

The HiddenBench transformation audit also bounds the benchmark claim. The transformed schema preserves the official shared and hidden fields and enforces split-evidence partitions, while source annotations leave hidden-profile minimality and shared-only decoy pressure partially recoverable. The benchmark should therefore be read as a structured split-evidence stress test with partially recoverable hidden-profile rationale evidence.

Targeted false specialists illustrate a second boundary: semantic attacks produce target steering with slice-dependent utility loss rather than universal collapse. Visible verifier filtering is also limited because it inspects commitments rather than raw KV state, producing honest false positives and missed hidden-state manipulations. Magnitude quarantine supports a scoped mitigation claim for evaluated nonsemantic anomalies; its threshold is post hoc, the reject cap must match the number of compromised specialists, and each rejection removes evidence.

The adaptive stress test is the sharpest boundary on quarantine. All evaluated norm-matched constructions evade the magnitude feature, and the white-box gradient-optimized variant is both evasive and damaging (Figure 2). Cheap fixed statistics and low-FP learned thresholding establish the detection ceiling. A learned top-1 sanitizer gives a useful 12-record ranking diagnostic, while the full-65 calibrated follow-up gives the decisive result: learned reject-1 trails a matched random-budget control under detector-aware attack and rejects 0/65 malicious specialists. The catastrophic case currently requires per-victim white-box gradient access, but the defense gap inside the honest magnitude band remains open.

The provenance-gated result has a precise boundary. The transport-layer HMAC-SHA256 manifest detects on-path post-handoff substitution of already-produced KV under an uncompromised key, after which dropping failed-provenance hidden state is stronger than visible downgrade on HiddenBench while the 100-record HotPotQA slice slightly favors visible downgrade. The dramatic endpoint/specialist attacks, including the white-box break, are recovery references for this policy analysis rather than attacks the MAC itself detects. Historical runs stored metadata without full tensor bytes, so replay binds recorded clean-vs-tampered identity plus metadata; byte-level detection is supported by the implementation and tensor self-tests. Remote attestation, production serving measurement, key-compromise protection, compromised endpoints signing malicious in-band KV, and semantic specialist behavior fall outside this transport-integrity mechanism.

Privacy functions as literature-motivated risk framing here. The measurements focus on integrity rather than leakage, attacks on private evidence, or privacy mitigations, so leakage remains a literature-grounded design concern rather than a measured result.

Future evidence should target compact bridges that approach text without hundreds of megabytes of KV transfer, robustness across decoding conditions and stronger semantic/adaptive attacks, and latent-state methods that complement MAC-backed transport integrity inside the honest magnitude band. The resulting scope is a bounded empirical audit of high-bandwidth latent memory under malicious specialists; compactness, privacy, and robust-security claims require additional evidence.

\section{Conclusion}
\label{sec:conclusion}

This paper provides a bounded audit of transported full-KV latent memory. The core systems-security question is empirical KV-transport integrity: when specialists pass opaque hidden-state objects to a coordinator, can visible commitments and latent-state diagnostics tell whether those objects remain trustworthy for split-evidence coordination? In the primary Qwen3-4B transformed HiddenBench setting, latent memory showed a clean split-evidence advantage over matched text collaboration, with statistical support strongest for F1. Qwen3-8B HiddenBench, full transformed HotPotQA, a reduced HotPotQA white-box/provenance slice, and full-HiddenBench influence mapping strengthen the evidence that the full-KV channel can carry useful evidence and reproduce the evaluated failure/defense pattern. Compactness and general superiority remain separate mechanism claims.

The security result is sharper. Once a specialist is compromised, the same transported hidden object creates an attack surface that visible text does not faithfully model. Semantic false-specialist behavior produced target steering and weaker, slice-dependent degradation, and the attacked influence diagnostic shows that such steering can be causally mediated by malicious specialist KV memory. Nonsemantic hidden-state manipulations were the more reproducible failure mode. The sign-flip contrast shows that destructiveness is attack-specific. The supported security claim is specific: interventions in the latent-thought state can dominate the coordinator when they interact with the full-KV channel, even when the visible commitment remains plausible.

The defense results argue for latent-state-aware auditing within clearly defined limits. Visible verifier filtering missed hidden corruptions and often rejected useful honest evidence. Magnitude quarantine recovered the evaluated one-malicious nonsemantic attacks and replicated on Qwen3-8B and 7,405-record HotPotQA diagnostics, while norm-matched attacks and a full-65 detector-aware follow-up showed why transport integrity must carry the defense claim. The stronger positive direction is MAC-backed transport-layer KV integrity. An HMAC-SHA256 manifest gate accepts 774/774 honest replayed payloads and rejects 295/295 recorded tampered payloads; after provenance failure, fail-closed handling restores much of the attacked utility on HiddenBench and the HotPotQA white-box slice. These results support a concrete systems boundary for post-handoff KV substitution under an uncompromised transport key, while key compromise, endpoint compromise, remote attestation, production serving, and in-band semantic malicious specialists remain outside the protection boundary.

The broader implication is that high-bandwidth latent memory should be evaluated as both a capability and an integrity liability. Anomaly detection is brittle under adaptive attack; a MAC-backed transport boundary is the constructive integrity mechanism for post-handoff KV substitution. Progress still requires latent-state defenses, compact bridges, measured serving paths, and broader replication across models, benchmarks, decoding regimes, semantic attacks, and stronger black-box attacks. Visible commitments are useful audit hooks, but transported KV state needs its own integrity boundary.

\clearpage
\appendix
\section{Ethical Considerations}
\label{sec:ethical-considerations}

This work studies a dual-use attack surface in latent-memory multi-agent systems. The artifacts include malicious-specialist attack code and logs, hidden-state corruption and quarantine diagnostics, adaptive stress tests, detector evaluations, and a white-box gradient-optimized attack. These materials can help researchers reproduce the observed failure modes and evaluate state-aware defenses. They could also help an actor with implementation-level access to a latent-memory pipeline corrupt specialist KV state or search for perturbations that evade simple post-hoc filters. We therefore frame the contribution as a bounded defensive audit, with attack artifacts contextualized by reproduction scope, paired defenses, and staged release choices.

The experiments are limited to open-weights research models, benchmark data, deterministic recorded runs, and reproduction metadata. They do not test production services, deployed multi-agent systems, private user data, or third-party infrastructure. The strongest demonstrated break assumes white-box coordinator-gradient access; the transfer diagnostics bound, but do not eliminate, the risk from gradient-free attackers. These boundaries limit the operational claim and are important release context, especially for systems that expose or transmit hidden-state objects between components.

The defensive intent is to make latent-channel risk visible before high-bandwidth hidden-state collaboration is treated as a trustworthy coordination mechanism. The attack suite is paired with contrastive and partial defense diagnostics: visible verifier filtering is documented as a limited audit mechanism, magnitude-based latent-state quarantine is documented as a post-hoc partial mitigation for the evaluated nonadaptive attacks, and adaptive stress tests document where that quarantine fails. This pairing is intended to discourage overclaiming from visible commitments or simple thresholds and to provide concrete baselines for future learned or training-time defenses.

No deployed target was tested or identified as affected by these experiments, so there is no product-specific vulnerability disclosure target in the current evidence package. If follow-up work evaluates a real deployed latent-memory service, hosted agent platform, or private infrastructure, disclosure should occur before publication. Affected operators should be notified privately, exploit details should be withheld until remediation timelines are agreed, and release artifacts should be scoped to non-production reproductions.

\section{Open Science}
\label{sec:open-science}

The artifact package is organized around a claim-to-artifact index. For each reported result, the index points to result JSON files, reports, run directories, metric tables, figure inputs, reproduction commands, and environment notes. The reproduction manifest records command lines, model and dataset settings, artifact paths, hashes, local ROCm metadata, the local HotPotQA white-box/provenance run, and remote L40S provenance for the full-validation HotPotQA suite while omitting secrets. The package also includes the transport MAC helper, replay driver, decision logs, and envelope examples used for the HMAC-SHA256 KV-integrity result. The current paper build uses completed evidence artifacts under \texttt{evidence/}.

During review, we will provide an anonymized artifact archive or submission-system artifact link. The review package will include redistributable transformed benchmark artifacts, run scripts, analysis scripts, result summaries, reproduction commands, environment manifests, and venue build source. For resources that cannot be redistributed directly because of benchmark, model, dependency, or third-party-code licenses, the package will provide hashes, provenance, acquisition instructions, and aggregate outputs sufficient to evaluate the corresponding claim. The archive will omit credentials, private tokens, identifying local paths, and non-benchmark data.

The white-box gradient attack script requires controlled handling because it combines hidden-state write access with coordinator gradients. It is claim-relevant and should be available to reviewers together with the defense scripts, detector evaluations, exact benchmark scope, and warnings that distinguish this upper-bound research stress test from black-box prompt-level attacks. Public release of that script may be staged or paired with defensive context, while aggregate results, reproduction metadata, transport-MAC code, and non-sensitive diagnostics can be released more broadly.

Before submission and public release, we will complete the remaining artifact checks: benchmark, model, and dependency-license review; redistribution review for imported third-party artifacts such as LatentMAS; log inspection for private prompts, credentials, and non-benchmark data; removal of identifying paths or metadata; and verification that the review archive remains double-blind. Any component that cannot be made public will be documented in the artifact README with the reason, reviewer-access mechanism when permitted, and substitute aggregate evidence needed to evaluate the affected claim.

\end{document}

%% file: figures/fig_protocol.tex
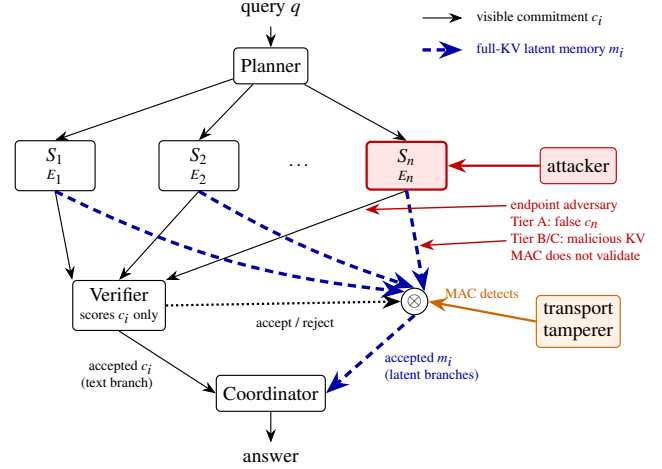
\begin{figure}[t]
\centering
\resizebox{\columnwidth}{!}{%
\begin{tikzpicture}[
  font=\scriptsize,
  box/.style={draw, rounded corners=1.5pt, align=center, inner sep=2.5pt, minimum height=4.5mm},
  spec/.style={box, minimum width=10mm},
  mal/.style={spec, draw=red!75!black, fill=red!10, thick},
  vis/.style={->, >=Stealth},
  lat/.style={->, >=Stealth, very thick, densely dashed, draw=blue!65!black},
  ctrl/.style={->, >=Stealth, densely dotted, thick},
  atk/.style={->, >=Stealth, thick, draw=red!75!black},
  atkptr/.style={->, >=Stealth, line width=0.5pt, draw=red!75!black},
  trans/.style={->, >=Stealth, thick, draw=orange!80!black},
]
\node (query) at (3.6,5.55) {query $q$};
\node[box] (planner) at (3.6,4.85) {Planner};
\node[spec] (s1) at (0.9,3.6) {$S_1$\\[-1.5pt]\tiny $E_1$};
\node[spec] (s2) at (2.7,3.6) {$S_2$\\[-1.5pt]\tiny $E_2$};
\node at (3.95,3.6) {$\cdots$};
\node[mal] (sn) at (5.3,3.6) {$S_n$\\[-1.5pt]\tiny $E_n$};
\node[box, draw=red!75!black, fill=red!10] (attacker) at (7.45,3.6) {attacker};
\node[box, draw=orange!80!black, fill=orange!10, align=center] (tamper) at (7.45,1.65) {transport\\tamperer};
\node[box] (verifier) at (1.7,1.85) {Verifier\\[-1.5pt]\tiny scores $c_i$ only};
\node[draw, circle, inner sep=0.8pt] (gate) at (5.4,1.9) {$\otimes$};
\node[box] (coord) at (3.6,0.75) {Coordinator};
\node (ans) at (3.6,-0.05) {answer};
\draw[vis] (5.5,5.45) -- (6.0,5.45);
\node[anchor=west, font=\tiny] at (6.05,5.45) {visible commitment $c_i$};
\draw[lat] (5.5,5.05) -- (6.0,5.05);
\node[anchor=west, font=\tiny, text=blue!65!black] at (6.05,5.05) {full-KV latent memory $m_i$};
\draw[vis] (query) -- (planner);
\draw[vis] (planner) -- (s1.north);
\draw[vis] (planner) -- (s2.north);
\draw[vis] (planner) -- (sn.north);
\draw[vis] (s1.south) -- (verifier.north west);
\draw[vis] (s2.south) -- (verifier.north);
\draw[vis] (sn.south) -- (verifier.north east);
\draw[lat] (s1.south) to[bend right=10] (gate.north west);
\draw[lat] (s2.south) to[bend right=6] (gate.north);
\draw[lat] (sn.south) -- (gate.north east);
\draw[ctrl] (verifier.east) -- node[below=1.5pt, pos=0.55, font=\tiny] {accept / reject} (gate.west);
\draw[vis] (verifier.south) -- node[pos=0.4, anchor=north east, inner sep=1.5pt, font=\tiny, align=right] {accepted $c_i$\\(text branch)} (coord.west);
\draw[lat] (gate.south) -- node[pos=0.4, anchor=north west, inner sep=1.5pt, font=\tiny, text=blue!65!black, align=left] {accepted $m_i$\\(latent branches)} (coord.east);
\draw[vis] (coord) -- (ans);
\draw[atk] (attacker.west) -- (sn.east);
\node[anchor=north, font=\tiny, text=red!75!black, align=left] (tiers) at (7.45,3.3)
  {endpoint adversary\\Tier A: false $c_n$\\Tier B/C: malicious KV\\MAC does not validate};
\draw[atkptr] (6.55,3.1) -- (4.78,3.05);
\draw[atkptr] (6.55,2.66) -- (5.46,2.6);
\draw[trans] (tamper.west) -- node[above=1pt, font=\tiny, text=orange!80!black] {MAC detects} (gate.east);
\end{tikzpicture}%
}
\caption{Role-sequenced protocol and threat model. Each specialist $S_i$ holds a private evidence partition $E_i$ and emits two channels: a short visible commitment $c_i$, which the verifier can score and filter, and high-bandwidth full-KV latent memory $m_i$ (prompt-token KV plus latent-thought KV), which reaches the coordinator without semantic inspection. A verifier rejection removes both $c_i$ and $m_i$ (filtered latent branch); the naive latent branch bypasses filtering, and the matched text branch replaces latent memory with visible commitments. A compromised endpoint/specialist can lie in the visible channel (Tier A) or generate malicious latent-thought KV (Tiers B/C); a transport MAC does not validate that endpoint-generated content. The MAC boundary instead covers an on-path tamperer that modifies already-produced KV after handoff.}
\label{fig:protocol}
\end{figure}

%% file: figures/fig_attack_defense.tex
\begin{figure}[t]
\centering
\begin{tikzpicture}
\begin{axis}[
  ybar,
  bar width=6pt,
  width=0.98\columnwidth,
  height=4.8cm,
  enlarge x limits=0.16,
  ymin=0, ymax=0.52,
  ytick={0,0.1,0.2,0.3,0.4},
  ylabel={Exact match},
  ylabel near ticks,
  ymajorgrids,
  major grid style={gray!25},
  axis x line*=bottom,
  axis y line*=left,
  symbolic x coords={TF,RAND,S8,GRAD},
  xtick=data,
  xticklabels={
    {targeted false\\(Tier A)\\quarantine n/a},
    {random corruption\\(Tier B)\\rejected 65/65},
    {scale-8\\(Tier B)\\rejected 65/65},
    {grad.\ norm-matched\\(Tier C)\\rejected 0/65}},
  x tick label style={font=\tiny, align=center},
  tick label style={font=\scriptsize},
  label style={font=\scriptsize},
  legend style={font=\tiny, draw=none, fill=none, at={(0.5,1.02)}, anchor=south, legend columns=-1, /tikz/every even column/.append style={column sep=5pt}},
  nodes near coords,
  every node near coord/.append style={font=\tiny, rotate=90, anchor=west, /pgf/number format/.cd, fixed, fixed zerofill, precision=3},
]
\addplot[ybar, fill=gray!35, draw=gray!60!black] coordinates {(TF,0.323) (RAND,0.323) (S8,0.323) (GRAD,0.323)};
\addplot[ybar, fill=red!60!black, draw=red!40!black] coordinates {(TF,0.231) (RAND,0.000) (S8,0.000) (GRAD,0.077)};
\addplot[ybar, fill=blue!45!black!55, draw=blue!50!black] coordinates {(RAND,0.338) (S8,0.385) (GRAD,0.077)};
\legend{clean (same-run), attacked, after quarantine}
\end{axis}
\end{tikzpicture}
\caption{Exact match on the full 65-record HiddenBench suite (Qwen3-4B, one compromised specialist) for the same-run clean latent branch, the attacked latent branch, and the accepted-specialist latent branch after post-hoc magnitude quarantine (reject at most one specialist whose post-key magnitude peer ratio exceeds 1.25). The quarantine rejects the naive nonsemantic attacks on 65/65 records with no honest false positives and restores their utility, but the white-box gradient-optimized norm-matched attack evades it on 65/65 records (0 rejections) and leaves the collapse uncorrected; the quarantine is not evaluated for the semantic targeted-false attack.}
\label{fig:attack-defense}
\end{figure}
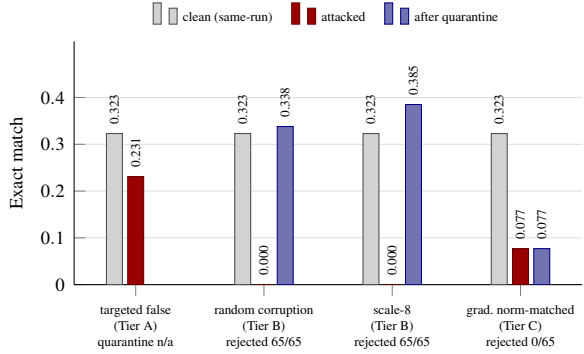